\documentstyle[12pt]{article}
\newcommand{\AmS}{{\protect\the\textfont2
  A\kern-.1667em\lower.5ex\hbox{M}\kern-.125emS}}

\begin{document}

\newcommand{\Otilde}{{O}\kern-.5em\lower 2.55ex\hbox{\Large$\tilde{}$}}
\newcommand{\Mtilde}{{M}\kern-.6em\lower 2.55ex\hbox{\Large$\tilde{}$}}


\vspace*{2cm}

\begin{center}
{\Large \bf Gauging the Shadow Sector with SO(3)}\\[1.5cm]
{\large \it R.J. Lindebaum, G.B. Tupper and R.D. Viollier}\\[.2cm]
Institute of Theoretical Physics and Astrophysics,\\
Department of Physics, University of Cape Town, Rondebosch 7701,\\
South Africa.\\
{\bf e-mail: viollier@physci.uct.ac.za}
\end{center}

\vspace{2cm}

\begin{abstract}
\noindent
We examine the phenomenology of a low-energy extension of the Standard Model, 
based on the gauge group $SU(3) \otimes SU(2) \otimes
U(1) \otimes SO(3)$, with $SO(3)$ operating in the shadow sector. This model 
offers $\nu_{e} \rightarrow \nu_{s}$ and $\nu_{\mu} \rightarrow 
\nu_{\tau}$ oscillations as the solution of the solar and atmospheric neutrino 
problems. Moreover, it provides a neutral heavy shadow lepton $X$ that could play 
the role of a cold dark matter particle. 
\end{abstract}

\newpage

With the accumulated evidence for neutrino oscillations [1] comes the 
challenge of understanding the origin of neutrino mass. Since the simple Higgs 
triplet [2] is ruled out by LEP [3], most approaches centre on singlet fermions 
and some variant of the see-saw mechanism [4]. The question then arises how 
these new particles fit into the larger theory.\\[.3cm]

One of the most novel proposals [5], inspired by $E_{8} \otimes 
E_{8}^{'}$ superstring theory, is that fermions, which are 
light because they are non-singlets under a low-energy 
``shadow gauge group'' $G^{'}$, could play an important role in this regard.
The implementation of this idea, however, involves several 
specific assumptions: $G^{'}$ is isomorphic to the Standard Model
$SU(3) \otimes SU(2) \otimes  U(1)$, with matter fields in the fundamental 
representation coming in three generations. None of these inputs 
are compulsory. In fact, in the superstring scenario,
compactification yields $E_{6}$ as the 
grand unified gauge group and it is also responsible for the 
generation structure, while the $E_{8}^{'}$ is left intact and can break down 
to $G^{'}$ in many ways. Furthermore, even if one can identify 
$SU(2)^{'} \times  U(1)^{'}$ as a subgroup of $G^{'}$, there are still many possible 
matter representations [6] including neutral heavy shadow leptons which could 
mix with the active neutrinos.\\[.3cm]

It is therefore important to explore other candidates for the low-energy group 
$G^{'}$, in order to obtain a clearer picture on this issue. Thus, in
this paper, we
will examine $G' = SO(3)$, the Georgi-Glashow 
model [7], which being vector like, is anomaly free. We assume no generation
replication and, since there is no information on the ``charged'' shadow
leptons, we 
dispense with the artifice of putting in a mass by hand. The shadow leptons 
appear as triplets
\begin{equation}
\vec{\psi}_{L} = \left(
\begin{array}{l}
E^{+}\\[.2cm]
X\\[.2cm]
e^{'-}\\ \end{array}
\right)_{L} \; \; \; \; \; \; , \; \; \; \; \; \;
\vec{\psi}_{R} = \left(
\begin{array}{l}
E^{+}\\[.2cm]
\nu^{'}\\[.2cm]
e^{'-}\\ \end{array}
\right)_{R}
\end{equation}
and the singlet $S_{R} = X_{R}$. At energies much below some large scale
$\Lambda$, 
a connection through the charge-neutral sector is phenomenologically 
afforded by
\begin{equation}
- {\cal{L}}_{G-G'} = \frac{f_{L \ell}}{\Lambda} \; \vec{\psi}_{L}^{T} \cdot 
\vec{\phi} \; C \; H_{2}^{\dag} \; L_{\ell} + \frac{f_{R \ell}}{\Lambda} \;
\bar{L}_{\ell} \; H_{2} \; \vec{\phi} \cdot \vec{\psi}_{R} + \mbox{h.c.} \; ,
\end{equation}
where $L_{\ell}$ and $H_{2}$ are the usual lepton doublets and u-like Higgs 
field, respectively, $\vec{\phi}$ is the shadow Higgs triplet and $\ell = e, \mu, \tau$ a 
generation index. Omitted in (2) is $X_{R}$, since by the survival hypothesis 
[8], it is expected to pick up a large Majorana mass.\\[.2cm]
At low energies with $< H_{2} > = v/\sqrt{2}, \; < \phi > = v'\;$ and
\begin{equation}
\vec{m}_{L/R} = \frac{vv'}{\sqrt{2} \Lambda} \; (f_{L/R \; e} \; , \;
f_{L/R \; \mu} \; , \; f_{L/R \; \tau})^{T} \; ,
\end{equation}
we obtain, in the basis ($\nu_{e}, \nu_{\mu}, \nu_{\tau}, \nu^{'c}, X_{L}$), the 
neutral mass matrix\\[.2cm]
\begin{equation}
\cal{\Mtilde} \hspace{.5cm} = \; \; \left(
\begin{array}{ccc}
{\Otilde} \; & \; \vec{m}_{R} \; & \; \vec{m}_{L}\\[.3cm]
\vec{m}_{R}^{T} \; & \; 0 \; & \; 0\\[.3cm]
\vec{m}_{L}^{T} \; & \; 0 \; & M\\ \end{array}
\right) \; ,
\end{equation}
where $M$ is the Majorana mass of $X_{L}$ resulting from the shadow see-saw
mechanism. 
This matrix has rank 1, the normalized zero eigenvector being
\begin{equation}
V^{0} \; = \; \frac{1}{| \sin \; \chi|} \left(
\begin{array}{c}
\hat{m}_{L} \; \times \; \hat{m}_{R}\\[.3cm]
0\\[.3cm]
0\\ 
\end{array} \right) \; ,
\end{equation}
and $\chi$ is the angle between $\hat{m}_{L}$ and $\hat{m}_{R}$. Taking
$M \gg |\vec{m}_{L/R}|, \cos \chi \ll 1$, and denoting
$x \equiv |\vec{m}_{L}|/m_{X}$, the
remaining eigenvalues and eigenvectors are
\begin{equation}
- \; m_{3} \approx - \vec{m}_{L}^{2}/m_{X} \; , \; \; V^{3} \simeq
\frac{1}{|\sin \chi| \sqrt{1 + x^{2} \sin^{2} \chi}} \;
\left(
\begin{array}{c}
\hat{m}_{R} \times \hat{m}_{L} \times \hat{m}_{R}\\[.3cm]
0\\[.3cm]
- x \; \sin^{2} \; \chi\\ 
\end{array} \right)
\end{equation}

\vspace{.3cm}

\begin{equation}
\pm \; m_{1,2} \; \simeq \; \pm \; |\vec{m}_{R}| \; + \;
\frac{m_{3}}{2} \; \cos^{2} \; \chi , \hspace{1cm} V^{1,2} \simeq
\frac{1}{\sqrt{2}} \;
\left(
\begin{array}{c}
\pm \; \hat{m}_{R}\\[.3cm]
1\\[.3cm]
0\\  \end{array} \right)
\end{equation}

\vspace{.3cm}

\begin{equation}
m_{4} \; \simeq \; M \; + \; \frac{\vec{m}_{L}^{2}}{M} \; \equiv \; m_{X} , 
\hspace{1.25cm} V^{4} \; 
\simeq \frac{1}{\sqrt{1+x^{2}}} \;
\left(
\begin{array}{c}
x \; \hat{m}_{L}\\[.3cm]
0\\[.3cm]
1\\  \end{array} \right) \; .
\end{equation}

Note that $| \Delta \; m_{12}| = | \Delta \; m_{03}| \cos^{2} \; 
\chi$.

\vspace{.3cm}

Since all oscillation solutions of the solar neutrino problem involve smaller
$|\Delta m|$ than the atmospheric neutrino problem, we take
$\hat{m}_{R}$ = (1, 0, 0), i.e. $\nu_{e} \rightarrow \nu_{s}$ with maximal
mixing, 
which is consistent with both the solar neutrino data and the nucleosynthesis bound 
[9]. As the atmospheric neutrino data are consistent with maximal mixing
$\nu_{\mu} \rightarrow \nu_{\tau}$, we take $\hat{m}_{L} = (\cos \chi,
\sin \chi/\sqrt{2}, \sin \chi/\sqrt{2})$, so that $\Delta m^{2}_{\rm atm}
\approx 3 \times 10^{-3}$ eV$^{2}$ fixes $m_{3} \approx 5.5 \times 10^{-2}$ eV.
The vacuum solution $\Delta m_{\odot}^{2} \approx$ 6.5 $\times$ 10$^{-{11}}$ 
eV$^{2}$ is consistent with $\chi \approx$ 90$^{o}$.\\[.2cm] 

Next we turn to cosmology in the presence of the neutral heavy lepton 
$X$. The decay $X \rightarrow \nu' \gamma'$, where $\gamma'$ is the shadow
photon, is absent because, $e'^{-}$ and $E^{+}$ being degenerate, their 
contributions [6] cancel. The $\nu' 2 \gamma'$ mode is allowed and can be 
estimated using the results of [10]. Taking $m_{E} \approx v'$ in the shadow 
see-saw mechanism, we arrive at

\begin{equation}
\tau_{X \rightarrow \nu' 2 \gamma'} \approx 10^{18} 
\left( \frac{\mbox{keV}}{m_{X}} \right)^{5} \;
\left( \frac{\Lambda}{v} \frac{e}{e'} \right)^{4} \; \mbox{yr} \; .
\end{equation}

A far more stringent constraint follows from the fact that
$X \rightarrow 3 \nu$ occurs via the weak neutral current for which a simple 
calculation yields

\begin{equation}
\tau_{X \rightarrow 3 \nu} = 1.7 \times 10^{16}
\left( \frac{\mbox{keV}}{m_{X}} \right)^{4} \; \mbox{yr} \; .
\end{equation}

Thus a neutral heavy shadow lepton of mass $m_{X} <$ 36 keV is stable on the 
lifetime of the universe,  and it can serve as dark matter if it is much less
abundant than ordinary neutrinos, so that it is consistent with the 
cosmological bounds.\\[.2cm] 

In order to avoid gross conflicts, in particular the production of $SO(3)$
monopoles in the early universe [11], it is 
assumed that the Big Bang explodes asymmetrically into the visible sector,
with the shadow sector remaining cold or empty [12]. Under this condition,
sterile neutrals such as $X$ may only be produced through 
oscillations [13] or gravitational interaction. If there is a large lepton
number asymmetry suppressing oscillations, these neutral leptons will
be produced with $\Omega \approx$ 1 and behave like cold dark matter,
for $m_{X}$
of the order of 10 keV [14].\\[.2cm]

Even under conservative assumptions [15], the $X$ particles
may form compact objects of a size $R \geq$ (100 km s$^{-1}/\sigma$) AU,
where $\sigma$ is the 
velocity dispersion. However, as has recently been shown the fermions
will undergo a first-order
gravitational phase transition [16], forming degenerate fermion stars as they
cool. The latent heat may be disposed of by ejecting 
some of the fermions or gravitational cooling [17].\\[.2cm]

There is circumstantial
astrophysical 
evidence for the existence of such a neutral fermion in the mass range of 10 to 20 
keV. In fact, modelling 
the violent compact dark object at the centre of M87 [18] with mass $M =$
(3.2 $\pm$ 0.9) $\times$ 10$^{9}$ $M_{\odot}$, as a degenerate 
fermion star near the Oppenheimer--Volkoff limit, constrains [19]
the fermion mass to
\begin{equation}
\mbox{12.4 keV} \; \; \leq \; m_{X} \; \leq \; \mbox{16.5 keV}   \; .
\end{equation}
Such a compact dark object would have a radius of 4.45 Schwarzschild radii. 
Thus there is little difference between a supermassive black hole and
a fermion star of the same mass, at the Oppenheimer-Volkoff limit, as the last
stable orbit around a black hole is 3 Schwarzschild radii anyway.\\[.2cm]
Similarly, modelling SgrA$^{*}$ near the centre of our galaxy [20] with mass
$M$ = (2.6 $\pm$ 0.2) $\times$ 10$^{6}$ M$_{\odot}$ as a fermion star, we obtain
a lower bound for the fermion mass of $m_{X} >$ 15.9 keV, from the observed 
motion of stars near SgrA$^{*}$ [21] which constrains the radius of the 
fermion star to less than 0.018 pc. 
The enigmatic radio and infrared emission of this 
object, interpreted in terms of standard thin disk accretion theory,
gives us an upper limit for the fermion mass $m_{X} < 18$ keV [22]
from the drop of the emission spectrum at infrared wavelengths.
Such a fermion star would differ very much from a supermassive black hole of 
the same mass, as the escape velocity from the fermion star would be only about 
$v_{\infty} \approx$ 1700 km/s. Virtually all supermassive compact dark 
objects that have been observed so far at the centres of galaxies have masses 
in the range of 10$^{6.5}$ to 10$^{9.5} M_{\odot}$.\\[.2cm]

Fixing $m_{X}$ = 16 keV, we obtain $|\vec{m}_{L}| \approx$ 30 eV, while 
$|\vec{m}_{R}|$ must be an order of magnitude smaller to allow $\nu_{e}$ and 
$\nu_{s}$ to serve as a hot dark matter component [23]. The scale appearing in 
eq.(2) as well as $v'$ can be estimated by combining eqs.(3), (6a) and $m_{E} 
\approx v'$; taking $|\vec{f}_{L}| \approx$ 1 we obtain
\begin{equation}
\Lambda \approx  \frac{v^{2}}{2 m_{3}} \approx 5.7 \times 10^{14} \; \mbox{GeV}
\end{equation}

\begin{equation}
\frac{v'}{v} \approx \sqrt{\frac{m_{X}}{2 m_{3}} } \approx 380 \; \; .
\end{equation}

\vspace{.3cm}

While $\Lambda$ is much below the Planck scale the identification of the 
$SO(3)$ as a low-energy subgroup of $E'_{8}$ is still possible as the string 
scale may be brought down by large compact dimensions [24].\\[.2cm]

In summary, we have examined a model with a low-energy shadow gauge group 
$SO(3)$ connected to the Standard Model through their neutral sectors. This 
model is capable of describing all the existing neutrino oscillation data
(except LSND), and 
it provides for cold dark matter in the form of a neutral heavy lepton $X$
in the shadow sector. This particle could play an important role in understanding the 
supermassive compact dark objects at the centres of galaxies in terms of 
degenerate heavy lepton stars.

\end{document}